\documentclass[aps,pra,preprint,superscriptaddress]{revtex4-1}

\usepackage{bm}
\usepackage{graphicx} 
\usepackage{color}
\usepackage{amsmath}
\usepackage{booktabs}

\begin{document}
 
\bibliographystyle{apsrev}

\title{Quantum Defect Theory description of weakly bound levels and Feshbach resonances in LiRb}

\author{Jes\'{u}s P\'{e}rez-R\'{i}os \footnote{Electronic mail:
jperezri@purdue.edu}}

\affiliation{Department of Physics and Astronomy, 
Purdue University,  47907 West Lafayette, IN, USA}

\author{Sourav Dutta}

\affiliation{Raman Research Institute, C. V. Raman Avenue, 
Sadashivanagar, Bangalore - 560080, India}

\author{Yong P. Chen}

\affiliation{Department of Physics and Astronomy, 
Purdue University,  47907 West Lafayette, IN, USA}

\affiliation{School of Electrical and Computer Engineering, 
Purdue University,  47907 West Lafayette, IN, USA}

\author{Chris H. Greene}

\affiliation{Department of Physics and Astronomy, 
Purdue University,  47907 West Lafayette, IN, USA}

\date{\today}

\begin{abstract}

 The multichannel quantum defect theory (MQDT) in combination with the 
 frame transformation (FT) approach is applied to model the Fano-Feshbach 
 resonances measured for $^{7}$Li$^{87}$Rb and $^{6}$Li$^{87}$Rb 
 [Marzok {\it et al.} Phys. Rev. A {\bf 79} 012717 (2009)]. The MQDT results 
 show a level of accuracy comparable to that of previous models 
 based on direct, fully numerical solutions of the the coupled channel
 Schr\"odinger equations (CC). Here, energy levels deduced from 
 2-photon photoassociation spectra for $^{7}$Li$^{85}$Rb are assigned 
 by applying the MQDT approach, obtaining  the bound state energies 
 for the coupled channel problem. Our results 
confirm that MQDT yields a compact description of photoassociation 
observables as well as the Fano-Feshbach resonance positions and widths.

\end{abstract}

\maketitle

\section{Introduction}

Ultracold molecules are currently generating tremendous interest in the
 atomic, molecular, and optical physics community (AMO) due to 
their potential applications as valuable tests and extensions of our 
understanding of processes in chemical physics, few-body physics 
and fundamental physics. In particular, ultracold molecules are expected
 to enable precise control of chemical reactions\cite{Krems-2008,Krems-2005,Hudson-2006}, 
 studies of novel quantum 
phase transitions\cite{Yi-2000,Goral-2002,Santos-2003}, 
realizations of novel dynamics in low-energy collisions\cite{Avdeenkov-2003}, 
and tests of the possible time variation of the fundamental constants of 
nature\cite{Hudson-2006-bis,Chin-2009}. 
Moreover, ultracold molecules could shed light on the fundamental laws 
and symmetries of nature, through measurements of the electric dipole moment 
of the electron\cite{Hudson-2002,Hudson-2011, ACME-2014}. 
These measurements have already been able to rule out some 
theories that were proposed as alternatives to the Standard Model.
 
Molecules can be brought down to the ultracold regime by either direct or 
indirect methods. Direct schemes employ external fields (electric 
fields for polar molecules, and magnetic fields for paramagnetic 
molecules), or sympathetic cooling via collisions with colder atoms
 that act as a dissipative medium for the molecules to move
 through. On the contrary, indirect methods 
start with an ensemble of ultracold atoms, and then external 
fields are used to glue the atoms together to form ultracold
 molecules. External magnetic field ramps have been used to create 
ultracold molecules by making use of Fano-Feshbach resonances
 associated with the atom-atom interaction\cite{Chin-2009}, 
in the so-called magnetoassociation (MA) technique\cite{Kohler-2006}. Laser
 fields can also provide useful interactions 
with ultracold molecules. A photon resonant
 with an excited atomic state can be absorbed while an ultracold 
atom collides with a ground state atom, in a  photoassociation
 (PA) process\cite{Jones-2006}.  After the absorption, the 
ultracold molecule in an excited state can decay to the ground state
 by spontaneous emission. 

The effectiveness of indirect cooling techniques depends on 
details of the atom-atom interaction, since those techniques are 
based on the existence of resonances. For this reason, indirect 
cooling methods can be a useful probe of the atom-atom interaction 
potential. Indeed, the results coming from MA or PA can be used 
to calculate an accurate atom-atom interaction through quantum 
scattering theory. This theory is based on the numerical solution 
of the radial coupled Schr\"odinger equations out to a large distance 
where the asymptotic conditions are applied\cite{Mott-1965}. While 
accurate, this method can be computationally demanding due to the
 large number of channels that are frequently involved, and because
 the scattering wave function requires propagation out to such long 
distances. In this respect, multichannel quantum
 defect theory (MQDT) can be an efficient alternative.

MQDT was born in atomic physics long ago, as a highly successful 
theory to explain the spectra of autoionizing states in complex atoms
 and the link between bound and continuum states of an outermost
 atomic electron\cite{Schrodinger-1921,Seaton-1966,Fano-1986}. 
Since those early developments, MQDT has been extended beyond 
the long-range Coulomb interaction to other long range 
potentials\cite{Greene-1979,Greene-1982}.  In particular, it has been applied
 to conventional atomic collisions\cite{Mies-1984,Fourre-1994}, 
and ultracold atomic collision\cite{Burke-1998,Mies-2000,Raoult-2004,Croft-2011,
Ruzic-2013,Pires-2014}. MQDT exploits the fact that at long-range 
the coupling between the channels is negligible, and this permits a
 systematic separation of short-range and long-range influences on
 the two-body physics. Specifically, for some long range potentials, an 
analytic solution of the scattering wave function can be found in 
terms of quantum defects that are almost energy independent. For 
other potentials it is advantageous to implement a numerical version 
of the long range QDT solutions, appropriately characterized in a way
 that makes the energy- and field-dependences of scattering observables
 as explicit as possible.

In some applications, MQDT is employed in an essentially exact
 manner, in that accurate solutions of the close-coupling equations 
are obtained out to a distance around $r_0=30-50$ a.u., and then 
matched to linear combinations of single channel solutions $(f_i,g_i)$ in
 the appropriate long range potential for each channel $i$.  In the present 
 context, of course, those are van der Waals long range potentials in 
 every channel.  For other applications, a simpler ``frame transformation 
 approximation™" that we abbreviate as MQDT-FT is utilized, as an 
 alternative to explicitly solving the coupled differential equations in 
 the inner region $r<r_0$.

The approximate MQDT-FT treatment is the version utilized
for the present study. The concept of the frame transformation
formulation is to start from single-channel values of the singlet
and triplet $s$-wave scattering lengths, which include no Zeeman
or hyperfine couplings. These give the phases of the wavefunction
 in those short-range scattering eigenchannels, and they can then be
 rotated through a unitary transformation matrix into the asymptotic 
 representation in which the atomic energy levels have been 
 diagonalized (with the internal and external magnetic couplings 
 included).  In some systems, accurate or approximate scattering 
 lengths $a_S$, $a_T$ are already known for the singlet and triplet
symmetries of an alkali metal dimer, respectively. The phase information
 contained in those scattering lengths can be 
recast as two short-range eigen-quantum-defects, $\mu_S,\mu_T$ which
 represent energy-analytic phaseshifts relative to the van der Waals $(f,g)$, 
 and which vary far more slowly with energy than the scattering lengths themselves. 

After frame-transforming these short range eigen-quantum-defects into the 
hyperfine plus Zeeman representation, a full $N \times N$ smooth reaction 
matrix is obtained for the system, and after closed channel elimination, 
Fano-Feshbach resonances emerge at various energies and magnetic field
 strengths $B$. (The closed-channel elimination step simply imposes correct 
 exponentially decaying boundary conditions in the energetically closed 
 channels.) The present study adopts the conventions for single-channel 
 long-range field solutions are chosen to be those introduced by 
 Ruzic {\it et al.}\cite{Ruzic-2013}. They represent a particular standardization of the 
long-range $(f,g)$, and there are four ``long-range QDT parameters" which 
are standard and reasonably simple in their energy dependence, and which 
embody the crucial energy-dependences and magnetic field dependences 
that are controlled by the van der Waals physics and the hyperfine plus 
Zeeman Hamiltonian. (There are minor differences between the 
standardizations introduced by Ref.\cite{Ruzic-2013} and those 
used in alternative variants of QDT ({\it e.g.} different from those of
Burke {\it et al.}\cite{Burke-1998} of Gao\cite{Gao-1998} or of 
Mies and Raoult\cite{Mies-2000}). The version
used here for the simplified frame transformation procedure is taken from
Pires {\it et al.}\cite{Pires-2014}. Our study here determines the short-range 
singlet and triplet quantum defects for two isotopologues $^{6}$Li$^{87}$Rb 
and $^{7}$Li$^{87}$Rb. The optimum values of the short-range quantum defects 
are chosen to be those that describe most accurately the position of the observed
Fano-Feshbach resonances. In another application developed in the following, 
MQDT is applied to assign the lines observed in two-color PA spectra
for $^{7}$Li$^{85}$Rb. Finally, some concluding remarks will address the 
applicability of MQDT to spectroscopic processes in ultracold physics.

\section{Multichannel quantum defect theory: bound state calculations}

Details about multichannel quantum defect theory (MQDT) can be 
found elsewhere\cite{Aymar-1996,Burke-1998,Ruzic-2013,Pires-2014}. 
Here only a brief description of the main features of the MQDT approach and 
its application to the calculation of bound states 
with coupled channels is presented. 

For two-body collisions in the presence of an external magnetic field, the 
wave function can be expanded in the basis of $N$ hyperfine plus Zeeman states 
(channels) that include the centrifugal angular momentum $l_{i}$

\begin{equation}
\label{eq-1}
\Psi(R,\Omega)=\frac{1}{R}\sum_{i=1}^{N}\Phi(\Omega)\Psi_{i}(R),
\end{equation}

\noindent
where $\Omega$ represents all angular coordinates and spin degrees of 
freedom. Eq.(\ref{eq-1}) must be a solution of the Schr\"odinger equation, 
leading to a set of coupled radial equations

\begin{equation}
\label{eq-2}
\sum_{j=1}^{N}\left[ \left(-\frac{d^2}{dR^2}+\frac{l_{j}(l_{j}+1)}{R^2}\right)\delta_{ij}+V_{ij}(R)\right]
\Psi_{j}(R)=E_{i}\Psi_{i}(R).
\end{equation}

\noindent
The matrix $V_{ij}(R)$ accounts for the coupling between different
 channels due to the interaction potential between the colliding particles. 
$E_{i}$ denotes the available kinetic energy for the {\it i}-th channel and it 
is given by $E_{i} = E - E_{i}^{thres}$, where $E$ is the collision energy 
and $E_{i}^{thres}$ stands for the Zeeman energy of the $i$ channel.

All lengths are expressed in units of the characteristic length scale $\beta$ 
associated to the potential $V$, and all energies are in units of the 
corresponding characteristic length scale
 $E_{\beta}=\frac{\hbar^2}{2\mu\beta^2}$, where $\mu$ is the reduced 
mass. The long-range behavior of $V$ specifies $\beta$ and hence 
$E_{\beta}$. In particular, the long-range interaction between two S-state atoms 
(such as two alkali atoms) leads to an isotropic van der Waals 
interaction $V=-C_{6}/R^6$, and the characteristic length is given by 
$\beta=(2\mu C_{6}/\hbar^2)^{1/4}$, denoted the van der Waals
 length and the corresponding energy 
scale is called van der Waals energy. In some references the van der Waals 
length is defined as $\beta /2$\cite{Chin-2010}.

For most two-body collisions involving neutral species the long-range tail 
of the potential is dominated by the van der Waals interaction. In such systems, the 
channels become approximately uncoupled beyond a radius $R_{M}$. In general, 
Eq. (\ref{eq-2}) has $N$ independent solutions that satisfy the 
physical boundary conditions $\Psi_{i}=0$ at $R$ = 0. The $N$ solutions 
of Eq. (\ref{eq-2}) can be regarded as the column vectors of the $N \times N$ 
solution matrix $M$. Thus, matching $M$ to single-channel reference 
wave function (uncoupled) $\hat{f}$ and $\hat{g}$ in each channel beyond $R_{M}$ 
defines the short-range reaction matrix $K$\cite{Aymar-1996},

\begin{equation}
\label{eq-3}
M_{ij}(R)=\hat{f}_{i}(R)\delta_{ij}-\hat{g}_{i}(R)K_{ij}^{sr}.
\end{equation}

In particular, $\hat{f}_{i}(R)$ and $\hat{g}_{i}(R)$ are the regular and irregular 
solutions of the uncoupled Schr\"odinger equations in the 
long-range potential $V^{lr}$,

\begin{equation} 
\label{eq-4}
\left[-\frac{d^2}{dR^{2}} + \frac{l_{i}(l_{i}+1)}{R^2} + V_{i}^{lr}(R) - E_{i} \right]\left( \begin{array}{l}
         \hat{f}_{i}(R) \\
        \hat{g}_{i}(R) \end{array} \right) = 0.
\end{equation}

The matrix $K^{sr}$ encapsulates all the information
about the short-range physics and channel coupling, 
whereas the standardized (smooth, analytic in energy) 
reference wave functions $\hat{f}_{i}(R)$ 
and $\hat{g}_{i}(R)$ describe the 
long-range physics. $K^{sr}$ and the 
linearly independent reference wave functions 
contain all the information necessary to 
calculate the scattering observables, through 
the scattering matrix, $S$. The calculation 
of the $S$ matrix requires two linearly 
independent, energy-normalized wave
 functions for open channels, and the bound-state 
wave function in each closed channel. 
As is standard in QDT, four long-range 
quantum defect parameters suffice to convert 
the smooth, short range reaction matrix $K^{sr}$ into 
the physical $S$ matrix which depends strongly on energy 
and magnetic field strength. The present calculations are 
based on the standardization of the long-range QDT 
parameters defined in Ref.\cite{Ruzic-2013}. Two of the 
long range QDT parameters, namely $A$ and $\mathcal{G}$, 
are used to generate a Wronskian-preserving 
transformation between the reference wave 
functions and two energy-normalized 
wave functions $f_{i}(R)$ and $g_{i}(R)$ in the open channels

\begin{equation}
\label{eq-5}
\left( \begin{array}{l}
         f_{i}(R) \\
        g_{i}(R) \end{array} \right) =
\left( \begin{array}{cc}
         A^{1/2} & 0 \\
        A^{-1/2}\mathcal{G} & A^{-1/2} \end{array} \right)
\left( \begin{array}{l}
         \hat{f}_{i}(R) \\
        \hat{g}_{i}(R) \end{array} \right).
\end{equation}

\noindent
The other long range QDT parameter at positive 
channel energy, $\eta_{i}$, represents
 for the asymptotic phase-shift of the energy-normalized 
$f_{i}$ and $g_{i}$ relative to the spherical 
Bessel functions. Finally, $\gamma$ 
is the long range QDT parameter at negative energy that describes the 
phase-shift of the reference wave functions $\hat{f}_{i}$ 
and $\hat{g}_{i}$, relative to the exponentially growing and decay 
solutions asymptotically which characterize bound-state solutions.
 The formulas to calculate those 
long range QDT parameters are given elsewhere\cite{Ruzic-2013}.

The MQDT parameters translate $K^{sr}$ 
into observables. For a given collision 
energy $E$, some channels will be open
 whereas the remain will be closed. Both 
kind of channels are included in the 
$K^{sr}$ matrix, which can be partitioned in 
terms of the open channel ($P$) and close
 channels ($Q$) contributions as

\begin{equation}
\label{eq-6}
K^{sr}=\left( \begin{array}{cc}
         K_{PP}^{sr} &  K_{PQ}^{sr}\\
        K_{QP}^{sr} &K_{QQ}^{sr} \end{array} \right).
\end{equation}

\noindent
However, the presence of closed-channel
 components in the $K^{sr}$ will lead 
to unphysical solutions at large distances, 
due to the presence of exponentially 
growing terms. This problem is removed 
by means of the MQDT step referred to as the 
``elimination of closed channels"€™\cite{Burke-1998}, after 
which the physical K-matrix is obtained from the formula:

\begin{equation}
\label{eq-7}
K=K_{PP}^{sr}-K_{PQ}^{sr}\left( K_{QQ}^{sr} +\cot{\gamma} \right)^{-1}K_{QP}^{sr}.
\end{equation}

\noindent
This expression shows explicitly the potentially resonant influence of
closed-channel pathways. The resulting $K$ matrix
has dimensions $N_{P}\times N_{P}$, with $N_{P}$ the 
number of open channels at the given collision energy $E$. 
In particular, from Eq. (\ref{eq-7}) discrete bound states 
can be obtained as the roots of the following equation: 

\begin{equation} 
\label{eq-8}
det\left( K_{QQ}^{sr}+\cot{\gamma}\right)=0,
\end{equation}

\noindent
where $\cot{\gamma}$ represents a diagonal matrix in channel space whose elements 
are equal to the closed channel QDT parameter $\cot{\gamma}$.

\subsection{Frame transformation machinery}

MQDT assumes that the short-range reaction matrix
 $K^{sr}$ depends very weakly on energy. Therefore, 
it can be calculated at just a few energies and then be 
interpolated between these values. In some cases, 
a single evaluated $K^{sr}$ matrix for a single chosen 
energy (usually close to the threshold) at zero 
magnetic field can be utilized to describe the
 scattering observables over a wide range of 
energies and magnetic fields. 

Generally, in scattering problems there is a representation 
where the Hamiltonian is diagonal at short-range and 
another one where the same Hamiltonian is diagonal
 at long-range. This difference in representations because 
 the terms in the Hamiltonian that dominate at small distance 
 often fail to commute with the terms dominant at large distance. 
 The frame transformation (FT) technique
 relies on an energy independent unitary transformation between
 the two representations. The MQDT-FT technique has been successfully
 applied to ultracold atomic collisions in the presence of 
an external magnetic field\cite{Burke-1998,Burke-1999,Gao-2005}.  
We follow here the method employed in a very recent study of the Li-Cs 
heteronuclear system\cite{Pires-2014}.

At short-range, due to the dominant role of the exchange
 energy, the collisional eigenstates are represented as 
 $|\alpha \rangle\equiv |\left(s_{A}s_{B}\right)S\left(i_{A}i_{B}\right)I \ FM_{F}\rangle$,
$s_{i}$ denotes the electronic spin of the{\it i}-th atom, 
$i_{i}$ stands for the nuclear spin of the {\ i}-th atom, $F$ is the total 
angular momentum of the molecule and $M_{F}$ is its 
projection on the quantization axes. In this basis  the 
$K^{sr}$ matrix is diagonal and reads as:
 
\begin{equation}
\label{eq-9}
K_{\alpha \alpha'}^{sr}=\tan{\left(\pi \mu_{\alpha}\right)}\delta_{\alpha \alpha'},
\end{equation}

\noindent
where $\mu_{\alpha}$ denotes the short-range single-channel quantum defects for the 
singlet $\mu_{S}$ and triplet $\mu_{T}$ states, which are approximated throughout this study as being
energy independent and magnetic field independent.

For the long-range part of the Hamiltonian, the hyperfine plus Zeeman
 energy is the dominant term of the Hamiltonian, and 
hence the collisional channels will be represented 
in the basis of the hyperfine+Zeeman eigenstates
 $|i\rangle =|m_{A}z_{A},m_{B}z_{B}\rangle$, which 
are a linear combination of the basis set 
$|f_{A}m_{A},f_{B}m_{B}\rangle$, whose superposition 
coefficients are functions of the magnetic field. The MQDT-FT 
method utilizes the energy 
independent unitary transformation between the 
short-range basis set $|\alpha \rangle$ and the 
long-range basis set $|i\rangle$, which is given by standard 
angular momentum coefficients (Clebsch-Gordan 
and Wigner 9-$j$ coefficients) and the Breit-Rabi 
eigenvectors, and we denote these transformation matrix elements as
 $\langle z_{A}z_{b}|f_{A}f_{B} \rangle^{(m_{A},m_{B})}$,
{\it etc}., and they are computed as

\begin{eqnarray}
\label{eq-10}
U_{i,\alpha}=\sum_{f_{A}f_{B}f} \langle z_{A},z_{B} |f_{A},f_{B} \rangle^{\left(m_{A},m_{B}\right)} 
\langle f_{A}m_{A},f_{B}m_{B}|FM_{F}\rangle   \nonumber \\
\times \langle \left(s_{A}i_{A} \right)f_{A}\left(s_{B}i_{B} \right)f_{B} |\left(s_{A}s_{B}\right)S\left(i_{A}i_{B}\right)I \rangle^{(F)}.
\end{eqnarray}

\noindent
The short-range reaction matrix is approximated here as being 
exactly diagonal in the short-range basis set, whereas the 
scattering observables are defined in the long-range basis set 
(hyperfine + Zeeman). Angular momentum coupling  
theory ensures the existence of the unitary transformation matrix connecting these 
two representations via Eq.(\ref{eq-10}), and therefore the smooth, short-range 
reaction matrix is given to an excellent approximation by:

\begin{equation}
\label{eq-11}
K_{ij}^{sr}=\sum_{\alpha}U_{i,\alpha} K_{\alpha}^{sr}U_{\alpha,j}^{T}, 
\end{equation}

\noindent
where $T$ denotes the matrix transpose. Note that $l$, the quantum number associated with the 
centrifugal angular momentum does not appear in Eq. (\ref{eq-10}), 
therefore the FT does not involve couplings between 
the atomic degrees of freedom (spin, nuclear spin, angular momentum) 
and the collisional degree of freedom.  In systems where magnetic dipole-dipole 
or quadrupole interactions are important, it could be desirable to include off-diagonal
coupling terms in $l$, but those are often sufficiently weak that they can be treated 
perturbatively.  The short-range quantum defects $\mu^{sr}$ do depend on $l$, but 
most of that $l$-dependence is known analytically;  a small $l$-dependent correction 
can be applied as in Ref.\cite{Pires-2014}.

\section{Analyzing Feshbach resonances for L\lowercase{i}R\lowercase{b}}

The MQDT-FT approach as presented in the previous section
 is applied here to describe Fano-Feshbach resonances in
 LiRb. In particular, we will focus on $^{6}$Li$^{87}$Rb and 
$^{7}$Li$^{87}$Rb, two isotopic mixtures
for which Feshbach resonances have been 
experimentally observed\cite{Deh-2008,Marzok-2009}. 

The MQDT-FT has been implemented by using the long-range potentials 
reported in Ref.\cite{Marzok-2009}, most importantly the
 long-range $C_{6}$ coefficient is 2550 a.u. whereas the 
 $C_{8}$ is 2.3416$\times$$10^{5}$ a.u. Those values corresponds 
 to the model I of Ref.\cite{Marzok-2009}. The 
short-range physics is fully characterized by means
 of the field independent and energy independent
 quantum defects $\mu_{S}$ and $\mu_{T}$, through
 the short-range reaction matrix. These short-range 
 quantum defects are adjusted up to find an optimal 
 agreement between the predicated position and width 
 of the Fano-Feshbach resonances. The FT technique is
 used to transform the short-range reaction matrix
 (see Eq. (\ref{eq-11})) into the long-range basis
 (hyperfine + Zeeman states). Finally, four long range QDT 
 parameters in each channel that depend on the channel energy are 
needed for establishing a relationship 
between the short-range and long-range physics, where the
 asymptotic conditions are applied. The present study uses
  these parameters, denoted as $\mathcal{G},A,\eta$ and 
  $\gamma$, which have been determined once and for all
   for a pure van der Waals potential at long range 
$-C_{6}/R^{6}$\cite{Ruzic-2013,Pires-2014}. The long-range
 quantum defects are standard and can be used for any 
 alkali-alkali collision. They have been tabulated as 
 functions of a single dimensionless variable which is the product 
of the van der Waals length and the wave number
 $k$\cite{Ruzic-2013,Pires-2014}. Finally, by means of Eq. (\ref{eq-8}), 
the magnetic field locations of the low energy Fano-Feshbach resonances 
are calculated. This procedure yields the resonance as 
positions, as functions of the short-range quantum defects. The 
short-range quantum defects, $\mu_{S}$ and $\mu_{T}$
may be regarded as fitting parameters to predict the positions 
of all resonances observed experimentally.
 In addition, the MQDT-FT approach also enables 
the computation of scattering total and partial cross sections, through the 
very well-known relation between the $K$ 
matrix [Eq. (\ref{eq-6})] and the $S$ matrix (see {\it e.g.} 
Ref.\cite{Mott-1965}).

The MQDT-FT results for $^{7}$Li$^{87}$Rb and $^{6}$Li$^{87}$Rb, using 
the hyperfine constants reported in Ref.\cite{Arimondo-1977}, in 
comparison with the CC calculations from Ref.\cite{Marzok-2009} 
are shown in Table I. Parenthetically, the MQDT-FT calculation 
reported here neglects entirely the spin-spin 
and second-order spin-orbit interactions. The fitting of the short-range 
quantum defects ($\mu_{S}$ and $\mu_{T}$) is performed by 
taking into account the $s$-wave as well as the $p$-wave Fano-Feshbach 
resonances. For the fitting, three independent fitting parameters are 
employed \cite{Pires-2014}, these are small deviation from the initial 
short-range quantum defects coming from the long-range 
potential of the model I of Marzok {\it et al.}\cite{Marzok-2009}. 
For the MQDT calculation a collision energy
 of 8 $\mu$K has been assumed. The quality of the results are 
measured by means of the weighted rms deviation $\delta$B$^{rms}$
 on the resonance position, which is defined as

\begin{equation}
\label{eq-12}
\delta B^{rms}=\sqrt{\frac{ \sum_{i=1}^{N} \delta_{i}^{2} \delta B_{i}^{-2}/N }{\sum_{i=1}^{N}\delta B_{i}^{-2}}}.
\end{equation}
  
\noindent
The summation is performed over $N$ Fano-Feshbach resonances 
for a given isotopic mixture, $\delta B_{i}$ denotes
 the experimental uncertainty of the resonance positions
 and $\delta_{i}=B_{0}^{model}-B_{0}^{exp}$, where
 {\it model} stands for MQDT-FT and CC, whereas {\it exp} denotes 
the experimental resonance positions. Table I shows that the MQDT-FT 
approach gives agreement with the position of measured resonances 
comparable to that achieved in previous CC 
calculations\cite{Marzok-2009}, which are far more 
computationally demanding. Indeed the weighted rms deviation
is smaller for the MQDT-FT results than for CC ones. Positions 
of the Fano-Feshbach resonances can be inferred from the 
divergences of the computed scattering length 
versus the magnetic field. Figure 1 shows the scattering 
length for $^{7}$Li$|1,1\rangle$ + $^{87}$Rb$|1,1\rangle$, calculated 
by means of the MQDT-FT approach.

\begin{table}[t]
  \begin{tabular}{ l  c c c c c c c  c c c c  }
    \hline
    \hline
Open Channel & &  &B$^{exp}_{0}$ (G) & $\Delta$B$^{exp}$(G) & &B$^{MQDT}_{0}$(G) & $\Delta$B$^{MQDT}(G)$ & & B$^{CC}_{0}$(G) & $\Delta$B$^{CC}$ (G) & l \\ \hline
    $^{6}$Li$|\frac{1}{2},\frac{1}{2}\rangle$ + $^{87}$Rb$|1,1\rangle$ & &   & 882.02 &1.27 & & 882.75 & & & 882.42 & & 1   \\ 
& &   & 1066.92&10.62 & & 1067.05 &6.26 & & 882.42 & 7.4 & 0 \\ 
     $^{7}$Li$|1,1\rangle$ + $^{87}$Rb$|1,1\rangle$ & &   & 389.5 &0.9 & & 390.64 & & & 390.2 & &1  \\ 
& &   & 447.4&1.1 & & 446.08 & & & 445.6 &  &1\\ 
& &   & 565&6 & & 563.19 &7.8 & & 568.8 & 7.9 & 0 \\ 
& &   & 649&70 & & 653.09 &204 & & 650.6 & 175 &0\\ 
$\delta$B$^{rms}$ (G)& &  &  & & & 0.67  &  & & 1.02 & &   \\ 
    \hline
\hline
  \end{tabular}

\caption{Comparison of the resulting $^{6}$Li$^{87}$Rb
 and $^{7}$Li$^{87}$Rb Fano-Feshbach resonance positions
 from MQDT and CC methods to the observed resonances. 
 A van der Waals $C_{6}$=2550 a.u and a  quadrupole-quadrupole 
 interaction term $C_{8}$ = 2.3416$\times$$10^{5}$ a.u have been 
 employed for the MQDT calculations (see text for details).
The experimentally observed positions B$^{exp}_{0}$ and 
widths $\Delta$B$^{exp}$ are taken from Ref.\cite{Marzok-2009}.
 The resonance positions and widths calculated by the 
MQDT approach are denoted by B$^{MQDT}_{0}$ and
 $\Delta$B$^{MQDT}$, respectively. The resonances 
positions based on the CC approach B$^{CC}_{0}$ 
and widths $\Delta$B$^{CC}$ are taken from the
 model I of Ref.\cite{Marzok-2009}. The positions
 and widths of the resonances are given in Gauss. The 
 nature of the resonances is shown in the last column 
 of the table, $l=0$ and $l=1$ represent $s$-wave and 
 $p$-wave respectively.
The weighted rms deviation $\delta$B$^{rms}$ 
(see text for details) is also shown. }
\end{table}

The widths of the $s$-wave Feshbach resonances shown in Table I 
for the MQDT approach have been obtained by first 
calculating the scattering length as a function of the external 
magnetic field. Next, the scattering length is fitted by means of the function 
$a(B)=a_{bg}\left(1+\sum_{i=1}^{N}\Delta B_{i}/(B-B_{i}) \right)$. 
Here $B_{i}$ denotes the position of the 
Feshbach resonance,  $\Delta B_{i}$ represents the width of 
the resonance, and the background scattering length $a_{bg}$ is a global parameter for the
 fitting. The MQDT results for the width of the resonances 
associated with $^{7}$Li$|1,1\rangle$ + $^{87}$Rb$|1,1\rangle$ 
are shown in Table I. The widths of the resonances are larger than 
the reported experimental values. A similar trend is observed 
for the CC calculations. On the contrary, the $s$-wave 
resonance of $^{6}$Li$^{87}$Rb is in good agreement with the
experimental reported data\cite{Marzok-2009}.

 \begin{figure}[h]
\centering
	\includegraphics[width=4.0 in]{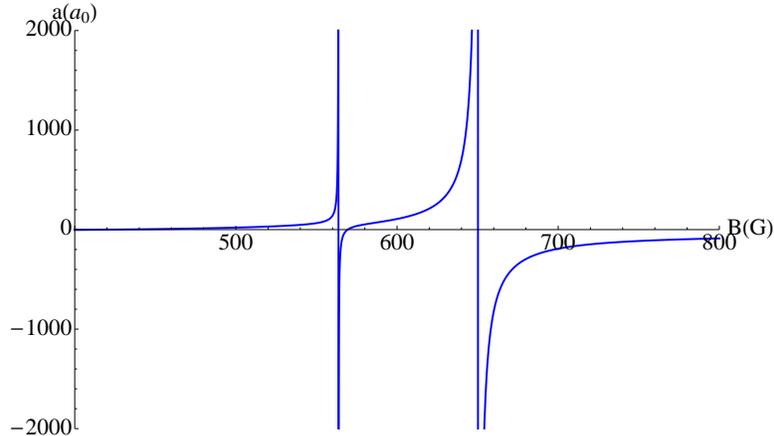}
	\caption{ Scattering length (in units of the Bohr 
radius) near the $s$-wave Fano-Feshbach resonances calculated by means of the
 MQDT-FT for  $^{7}$Li$|1,1\rangle$ + $^{87}$Rb$|1,1\rangle$, 
as a function of the magnetic field (in Gauss).}
\end{figure}

 The optimal short-range quantum defects, as well as the 
scattering lengths extracted from them, are shown in Table II. In 
addition, the $p$-wave short-range quantum defects are also 
shown in the same table for all the isotopologues for LiRb 
studied in this work. The scattering length calculated by the MQDT-FT 
compares well with the previously reported CC
 calculation\cite{Marzok-2009}, showing that MQDT 
 can accurately predict such Fano-Feshbach resonance 
 positions.

\begin{table}[h]
  \begin{tabular}{ l  c c c c c c c c c}
    \hline
    \hline
    & \multicolumn{4}{c}{$s$-wave} & & & \multicolumn{2}{c}{$p$-wave} \\
Molecule& $\mu_{S}$ & $\mu_{T}$  & $a_{S}$ (a$_{0}$) & $a_{T}$ (a$_{0}$) & && $\mu_{S}$ & $\mu_{T}$  \\ 
\cline{2-5}
\cline{7-9}
$^{6}$Li$^{87}$Rb & 0.2572 & 0.3184 & -1.87 & -22.70 & && 0.0045 & 0.0648 \\
$^{7}$Li$^{87}$Rb & -0.0817 &0.3845 & 53.20 & -68.85 &&&  -0.3282 & 0.1380 \\ 
$^{7}$Li$^{85}$Rb & -0.1259 & 0.3707 & 59.73 & -55.49 &&& -0.3724 & 0.1242  &    \\ 
    \hline
\hline
  \end{tabular}
\caption{Calculated $s$-wave and $p$-wave short-range quantum defects and scattering 
lengths (only for the $s$-wave) for the uncoupled singlet (S) and triplet (T) states of
 different isotopic mixtures of LiRb molecule. The scattering length
 is given in units of the Bohr radius (a$_{0}$).}
\end{table}

\section{Two-Photon Photoassociation: analysis of the least bound states of $^{7}$L\lowercase{i}
$^{85}$R\lowercase{b} }

Feshbach resonances have not been observed to date for the 
$^{7}$Li - $^{85}$Rb system. However, we have
recently measured the least bound states of $^{7}$Li $^{85}$Rb 
 using Raman-type two-photon PA, the
experimental details of which will be described elsewhere\cite{Dutta-2015}.

For shallow bound states, the associated wave functions 
mainly sample the long-range tail of the potential. In such
a system, the MQDT approach becomes a valuable tool for
the calculation of such bound states. For Li - Rb, ground 
state collisions can occur in any of the two distinct 
potentials, X$^{1}\Sigma^{+}$ and $a^{3}\Sigma_{+}$. 
Both potentials will be coupled due to the presence 
of hyperfine interaction in both atoms. The MQDT approach 
naturally includes such coupling between the singlet and 
triplet potentials through their respective quantum defects 
$\mu_{S}$ and $\mu_{T}$ [see Eq. (\ref{eq-9})] and of 
course the hyperfine plus Zeeman terms in the Hamiltonian
 which don'€™t commute with the total spin operators. For these 
calculations the hyperfine constants or Ref.\cite{Arimondo-1977} 
have been used, and for the calculation of the quantum 
defects, the singlet and triplet scattering lengths reported
 in Ref.\cite{Marzok-2009} have been utilized. The long-range
  coefficients of the previous section have been employed here 
  as well. The $B=0$ binding energies for the $s$ and $p$-wave 
  bound states calculated using the MQDT-FT approach are listed 
  in Table III. Those bound states have been obtained through the 
  short-range quantum defects listed in Table I. The quantum defects
   have been obtained from the calculated scattering length reported
    by Marzok {\it et al.}\cite{Marzok-2009}, and fitted a posteriori. In particular, 
we have employed the same fitting parameters that were obtained for the 
fitting of $^{7}$Li$^{87}$Rb, since the isotopic effect of Rb should be very 
small in comparison with the case at hand. In table III 
  it is shown the total $F$ associated to each state. The $F$ quantum number
   has been calculated by means a block diagonal procedure, {\it i.e.}, 
   by varying the number of channels taken into account for the 
   calculations of bound states. In each step, a new block of channels 
   associated to a given $F$ were included, and hence revealing the 
   nature of each bound state. 

\begin{table}[h]
  \begin{tabular}{ c  c c c c c c c c c c c cc }
    \hline
    \hline
 &$(f_{Li}$,$f_{Rb}$ ) & & \multicolumn{1}{c}{(1,2 )}   & \multicolumn{1}{c}{ (2,2)}  &\multicolumn{1}{c}{ (1,3)}  
 &\multicolumn{1}{c}{(2,3)} & & & \multicolumn{1}{c}{(1,2 )}   & \multicolumn{1}{c}{ (2,2)}  &\multicolumn{1}{c}{ (1,3)}  
 &\multicolumn{1}{c}{(2,3)}   \\ \hline
 & &  &\multicolumn{4}{c}{$s$-wave} & & & \multicolumn{4}{c}{$p$-wave} &  \\
\cline{4-7}
\cline{10-13}
& $F$  & & & & & & & &\\
&1  && -1.67 & -0.86 & 1.37 & 2.17 & & & -3.37 & -2.57& -0.34 & 0.46  \\
&2 &  &-1.58 & -0.77& 1.45 & 2.26& && -3.25 & -2.45 & -0.21 & 0.59  \\
&3  & &-1.44 & -0.64& 1.59& 2.40 & & & -3.00 & -2.20 & 0.03 & 0.83\\
&4  & & &-0.43&  1.80&2.61&&  & & -2.02 & 0.21 & 1.01\\

&5  &   &&& & 4.03& &  && &  & 3.38\\
&4  &&  &1.22& 3.46& 4.26 & &&& 0.66 & 2.89 & 3.70\\
&3  && 0.63  &1.43& 3.67 & 4.47 &&& 0.08 & 0.89 & 3.12 & 3.92 \\
&2  & &0.77  & 1.58& 3.81& 4.61 & && 0.24 & 1.05 & 3.28 & 4.08\\

&4  &  && 5.08& 7.31& 8.11 & && & 4.61 & 6.84 & 7.64\\
&3  && 4.40 & 5.21& 7.45& 8.25 &&& 3.94 & 4.75 & 6.98 & 7.78\\
&2  && 4.53& 5.34&   7.57&  8.37 &&& 4.08 & 4.88 & 7.11 & 7.92 \\
&1 & & 4.64&5.44& 7.68& 8.48 & && 4.18 & 4.98 & 7.22 & 8.02\\ 

&0  && 6.12& 6.93& 9.16&  9.96 & & & 5.61 & 6.42 & 8.65 & 9.45\\
&1  && 6.22&  7.03& 9.26&  10.06 & & & 5.71 & 6.52 & 8.75 & 9.55  \\
&2  && 6.40&  7.21&   9.44 & 10.24&& & 5.89 & 6.70 & 8.93 & 9.73\\
&3  && 6.64& 7.45&  9.68&  10.48&& & 6.13 & 6.93 & 9.17 &9.97\\ 

    \hline
\hline
  \end{tabular}
\caption{$s$-wave and $p$-wave bound states binding energies (in GHz) 
calculated using the MQDT-FT approach. In the first column the total $F$ for each bound 
state is shown (ignoring orbital angular momentum). The binding energies are referred 
to the threshold shown in the 
first row of the table.}
\end{table}

The scheme for Raman-type two-photon PA is shown in Figure 2. The
 ultracold Li and Rb atoms in a dual species magneto-optical trap 
 (MOT) collide predominantly in the Li (2s, $f_{Li}$ = 2) + 
 Rb (5s,$f_{Rb}$ = 2) channel\cite{Dutta-JPB,Dutta-EPL,Dutta-pra}. They are 
 photoassociated, using a PA laser at frequency $\nu_{PA}$, to 
 form weakly bound electronically excited LiRb$^{*}$ molecules
  in a rovibrational level denoted by $\nu'$ from which they spontaneously
decay to the electronic ground state or to free atoms leading to loss of 
atoms from the MOT \cite{Dutta-EPL,Dutta-pra}. This loss of atoms is detected as a 
decrease in the fluorescence of the Li MOT. A second laser, called the
 Raman laser with frequency $\nu_{R}$, is scanned across a 
 bound-bound $\nu' \leftrightarrow \nu''$ transition between the
  electronically excited and ground states. The polarization of the Raman 
  laser is perpendicular to the polarization of the PA laser. When the Raman laser
 is resonant with the $\nu' \leftrightarrow \nu''$ transition it causes 
 an Autler-Townes splitting in the $\nu'$ level leading to the PA 
 laser going out of resonance\cite{Abraham-1995}, hence suppressing
  the PA induced atom loss. This suppression of atom loss is a signature 
 of a Raman resonance and the binding energy of the $\nu''$ level is given by
 $\Delta \nu = \nu_{R}-\nu_{PA}$.

\begin{figure}[h]
\centering
	\includegraphics[width=4.0 in]{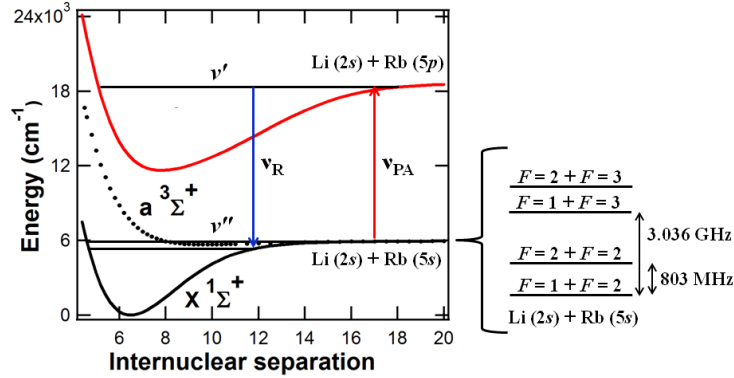}
	\caption{ Scheme used for Raman type 2-photon 
	spectroscopy for  $^{7}$Li$^{85}$Rb. The inset to the right shows the different hyperfine 
	channels along which the ground electronic state dissociates 
	asymptotically. See text for details.}
\end{figure}

In our experimental set up the linear polarizations of the lasers are perpendicular 
to each other, leading to a new set of selection rules. In particular, since 
Li and Rb are colliding predominantly through the $s$-wave initially, this implies that only 
$s$ and $d$-wave bound states will be allowed following the Raman process. For this reason, only 
the $s$-wave bound states have been considered for the assignment of 
the observed 2-photon PA lines, as are shown in Table IV. In the same 
table are shown the experimentally observed 2-photon PA lines up to
 5 GHz of binding energy.

 \begin{table}[h]
  \begin{tabular}{ p{1cm}p{0.8cm}p{2.5cm} p{2.5cm} p{2.5cm} p{2.5cm}  p{4cm} }
    \hline
    \hline
     & &($f_{Li},f_{Rb}$)& ($f_{Li},f_{Rb}$)& ($f_{Li},f_{Rb}$)& ($f_{Li},f_{Rb}$) &  \\ 
Name & & (2,2) &  (1,2) & (1,3) &  (2,3)& Assignment   \\ \hline
$\alpha$ & & -0.63 (-0.64)& -1.43 & 1.61 & 2.41 & ($f_{Li}$ = 2,$f_{Rb}$ = 2), $F$=3\\
$\beta$ & &-0.31 & -1.11 & 1.93 & 2.73 (2.61) &  ($f_{Li}$ = 2,$f_{Rb}$ = 3), $F$=4\\
$\gamma$  && 0.98 & 0.18 & 3.22 & 4.02 (4.03) &($f_{Li}$ = 2,$f_{Rb}$ = 3), $F$=5  \\
$\delta$ &&1.26 (1.22) & 0.46 & 3.5 & 4.3 & ($f_{Li}$ = 2,$f_{Rb}$ = 2), $F$=4 \\
$\epsilon$ && 1.55 (1.58) & 0.75 & 3.79 & 4.59 &($f_{Li}$ = 2,$f_{Rb}$ = 2), $F$=2 \\
$\xi$ && 3.2 & 2.4 & 5.44 & 6.24 &   \\
$\eta$&& 4.5 & 3.7 & 6.74 & 7.54  &  \\ 
    \hline
\hline
  \end{tabular}
\caption{Experimentally measured binding energies for s-wave bound states. The 
binding energies (in GHz) are represented in each column for each possible channel: 
Li (2s, $f_{Li}$ = 2) + Rb (5s, $f_{Rb}$ = 2), Li (2s, $f_{Li}$ = 1) + Rb (5s, $f_{Rb}$ = 2),
 Li (2s, $f_{Li}$ = 1) + Rb (5s, $f_{Rb}$ = 3) and Li (2s, $f_{Li}$ = 2) + Rb (5s, $f_{Rb}$ = 3), 
 from left to right, respectively. By comparing these energies with the MQDT 
 values shown in Table III, tentative assignments are made and shown in 
 the last column. The theoretical prediction for the bound
  states energies based on the MQDT approach are shown in parenthesis.  }
\end{table}

 There are two distinct potentials, X$^{1}\Sigma^{+}$ and $a^{3}\Sigma_{+}$, at small
 internuclear separations but at large internuclear separations 
 both potentials approach the Li (2s) + Rb (5s) asymptote with 
 the same $C_{6}$ coefficient. The bound states measured in 
 our experiments are very close to the dissociation limit where
the two potentials can be described with a single 
$C_{6}$ coefficient, and it is also the region for which the MQDT approach is expected 
 to give reliable results. Since Li (2s) and Rb (5s) atoms have hyperfine
structure, for $B=0$ the electronic potentials at very large internuclear separation
must dissociate along one of the four hyperfine channels: Li (2s, $f_{Li}$ = 1)
 + Rb (5s, $f_{Rb}$ = 2), Li (2s, $f_{Li}$ = 2) + Rb (5s, $f_{Rb}$ = 2), Li (2s, $f_{Li}$ = 1)
  + Rb (5s, $f_{Rb}$ = 3) or Li (2s, $f_{Li}$ = 2) + Rb (5s, $f_{Rb}$ = 3). Since we start 
 with atoms colliding along the Li (2s, $f_{Li}$ = 2) + Rb (5s, $f_{Rb}$ = 2) channel 
 all bound levels corresponding to this channel will have positive values 
 for $\Delta \nu$. The same is true for the potential dissociation along the 
 Li (2s, $f_{Li}$ = 1) + Rb (5s, $f_{Rb}$ = 2) hyperfine channel. Negative values
 $\Delta \nu$ necessarily have to be bound levels of the potentials 
 dissociating along the Li (2s, $f_{Li}$ = 1) + Rb (5s, $f_{Rb}$ = 3) or the 
 Li (2s, $f_{Li}$ = 2) + Rb (5s, $f_{Rb}$ = 3) hyperfine channels. In our case the 
 binding energies are measured with respect to the  Li (2s, $f_{Li}$ = 1) + Rb (5s, $f_{Rb}$ = 2)
channel, so the relevant atomic hyperfine energy needs to be added 
or subtracted in order to calculate the binding energy measured with 
respect to different channels. To calculate the binding energy with 
respect to  Li (2s, $f_{Li}$ = 1) + Rb (5s, $f_{Rb}$ = 2) channel we subtract
 0.803 GHz (the Li hyperfine splitting), to calculate the binding energy 
with respect to the Li (2s, $f_{Li}$ = 1) + Rb (5s, $f_{Rb}$ = 3) channel we 
add 3.04 GHz (the Rb hyperfine splitting) and to calculate the binding
 energy with respect to the channel  Li (2s, $f_{Li}$ = 2) + Rb (5s, $f_{Rb}$ = 3),
we add 3.843 GHz (sum of Li and Rb hyperfine splitting) to the 
observed values of $\Delta \nu$ (see table IV). 

Finally, some discussion in relation with the assignments of the 
observed levels is pertinent. The presented assignment shown 
in Table IV have been done by comparing the observed position 
of the peaks and the predicted bound state energies (Table III). However, 
the last two peaks have not been assigned to any $s$-wave bound 
state. Since these states are deeper that the previous one, they could 
be explore part of the interaction potential that needs to be described 
beyond the $C_{6}$ coefficient, and probably it would explain the 
discrepancies between our predictions and the observed peaks. 
Another reason would be that those states are associated to 
$d$-wave bound states, and these are beyond the approach presented.

 \section{Summary and Conclusions}
  
 The MQDT approach has been employed in two
  very different scenarios: Fano-Feshbach resonance 
description, and assignment of the 2-photon PA spectra. 
MQDT in addition with the frame transformation has been employed to fit 
the observed experimental positions of the Fano-Feshbach
 resonances for $^{7}$Li$^{87}$Rb and $^{6}$Li$^{87}$Rb. 
 The $s$-wave quantum defects
associated with the triplet and singlet potentials are used as the fitting
parameters. Then the scattering lengths for triplet and 
singlet potentials have been obtained through the 
obtained quantum defects. The resulting fit using 
MQDT turns out to be as accurate as one obtains by solving the 
coupled-channel Schr\"{o}dinger equation, but with 
much less numerical effort. 

For 2-photon PA spectroscopy, MQDT is an 
excellent tool for the assignments of the observed
 spectra. The capability of MQDT for calculating shallow bound 
 states (dominant by the long-range tail 
 of the interaction) between coupled electronic states has 
 been exploited. An outcome of this work is the assignments  of 
 our experimentally observed 2-photon PA lines. MQDT 
 may also used for calculating the scattering length 
 associated with the triplet and singlet electronic potentials, 
 similar to our analysis of the Fano-Feshbach resonances. 
 This is something that will be addressed in a subsequent 
 publication.

This work has been supported in part by the AFOSR-MURI 
program, and in part by NSF under grand number PHY-130690.  
We thank Brandon Ruzic for providing access to unpublished 
programs and data used in the MQDT calculations.

\bibliography{MQDT}

\end{document}